# Transverse oscillations in solar coronal loops induced by propagating Alfvénic pulses

Luca Del Zanna[1], Eveline Schaekens[1,2], Marco Velli[1]

[1] Dipartimento di Astronomia e Scienza dello Spazio, Università di Firenze, Largo E. Fermi 2, 50125 Firenze, Italy
e-mail: ldz@arcetri.astro.it

[2] Department of Astrophysics, University of Nijmegen, PO Box 9010, 6500 GL Nijmegen, The Netherlands

**Abstract.** The propagation and the evolution of Alfvénic pulses in the solar coronal arcades is investigated by means of MHD numerical simulations. Significant transverse oscillations in coronal loops, triggered by nearby flare events, are often measured in EUV lines and are generally interpreted as standing kink modes. However, the damping times of these oscillations are typically very short (from one to a few periods) and the physical mechanism responsible for the decay is still a matter of debate. Moreover, the majority of the observed cases actually appears to be better modeled by *propagating*, rather than standing, modes. Here we perform 2.5-D compressible MHD simulations of impulsively generated Alfvén waves propagating in a potential magnetic arcade (assumed as a simplified 2-D loop model), taking into account the stratification of the solar atmosphere with height from the photosphere to the corona. The results show a strong spreading of the initially localized pulses along the loop, due to the variations in the Alfvén velocity with height, and correspondingly an efficient damping of the amplitude of the oscillations. We believe that simple explanations based on the effects of wave propagation in highly inhomogeneous media may apply to the majority of the reported cases, and that variations of the background density and Alfvén speed along the loop should be considered as key ingredients in future models.

## 1. Introduction

Thanks to the unprecedented spatial and temporal resolution of the EUV instruments on board the SOHO (*Solar and Heliospheric Observatory*) and TRACE (*Transition Region and Coronal Explorer*) spacecrafts, there is a branch of Solar Physics which has been undergoing a very rapid development, namely *coronal seismology* (see Roberts 2000 for a review). The accurate observation of oscillations in different coronal structures (e.g. prominences, loops, plumes), together with density and temperature measurements, can be compared with theoretical models to infer plasma parameters otherwise difficult to measure, such as the coronal magnetic field strength. The basic theoretical framework needed to implement such a diagnostic tool is magnetohydrodynamics (MHD), which is assumed in most of the models for coronal equilibrium structures and wave propagation.

An important example of application of MHD coronal seismology is provided by loop oscillations. Aschwanden et al. (1999) and Nakariakov et al. (1999) have both analyzed TRACE observations of the motions induced by a flare (occurred on 1998 July 14) in a system of loops of a magnetic arcade. Nakariakov et al. (1999) investigated in particular the transverse oscillations of a single bright loop and measured their damping (*e*-folding) time. This turned out to be extremely short, just three times the oscillation period (approximately 15 minutes for a period of about 5 minutes). The result was interpreted by the authors as viscous damping of global kink standing modes in high density flux tubes, although the data could be fitted only by assuming (shear) Reynolds numbers of $10^5 - 10^6$, very low when compared to what is generally assumed for an almost collisionless corona: $R \sim 10^{12} - 10^{14}$ (similar values are supposed to hold for the magnetic Reynolds number). Such low values for the collisional viscosity and resistivity coefficients may be of paramount importance for the solution of the coronal heating problem (see e.g. Malara & Velli 1994; Del Zanna & Velli 2002, for reviews on wave-based heating mechanisms).

The above interpretation has obviously opened a lively debate in the solar community. Other observations have followed (Schrijver & Brown 2000; Nakariakov & Ofman 2001; Schrijver et al. 2002; Aschwanden et al. 2002; De Moortel et al. 2002) and alternative explanations for the quick damping have been put forward. Schrijver & Brown (2000) proposed that the observed oscillations are actually induced by photospheric footpoint motions, amplified in the corona if the original displacement occurs in the close vicinities of a magnetic null point. This scenario is supported by the fact that the wave period matches the typical photospheric sound wave period (around 5 minutes), although a time-dependent model aimed at explaining the damped oscillations is still missing. Other models based on the photosphere–corona connection consider the possible leakage of standing kink modes at the transition region density jump. Based on the analysis by Berghmans & De Bruyne (1995), Nakariakov et al. (1999) discarded that hypothesis because the predicted damping times were far too long (see however De Pontieu et al. 2001). A 1-D numerical model by Ofman (2002) later confirmed that the leakage damping time should be longer than the observed one by at least a factor five. The last group of theoretical interpretations still relies on the original modeling of the loop as a flux tube. If its boundaries are sharp enough, with correspondingly large Alfvén speed gradients, then dissipation is likely to be enhanced there, either by phase mixing or by resonant absorption (Roberts 2000; Ruderman & Roberts 2002; Ofman & Aschwanden 2002; Goossens et al. 2002; Aschwanden et al. 2003).

Other than just clear cases of standing modes, different kinds of transverse oscillations are observed in coronal loops in response to nearby flares. In particular, Aschwanden et al. (2002) analyzed TRACE observations of 26 oscillating loops and concluded that global modes are actually very rare, the majority of the events being better explained by impulsively generated MHD waves, which then propagate back and forth along the loop and rapidly decay. This scenario appears to be more realistic, since a flare is likely to affect the loop locally at the initial time while only later the disturbances will propagate. Also, in most cases asymmetric oscillations and irregular wave propagation periods are observed, as expected for a random location of the initial pulse along the loop. The authors find even smaller ($1-3$) ratios between decay times and wave periods (especially for small loops, say with total length less than $10^5$ km) and conclude by claiming that footpoint leakage may still explain the damping, provided the transition region temperature profile is not too sharp and the chromospheric scale height is large enough.

Finally, also compressive oscillations are observed in coronal loops by the SOHO and TRACE instruments. These can be either propagating waves along cool loops (Berghmans & Clette 1999; De Moortel et al. 2000; Robbrecht et al. 2001), or standing modes in hot post-flare loops (first seen in both Doppler shift and intensity by Wang et al. 2003) with rapid decay times similar to the case of kink modes. All these oscillations are modeled as slow magnetosonic longitudinal modes (Nakariakov et al. 2000), where thermal conduction is expected to play a major role in the damping mechanism (Ofman & Wang 2002; De Moortel & Hood 2003; Mendoza-Briceño et al. 2004). Impulsive generation of acoustic modes has also been recently modeled (Nakariakov et al. 2004; Tsiklauri et al. 2004).

While the various damping models have been separately analyzed as described above a set of full MHD multidimensional simulations is still missing. For example, all the theoretical models concerning partial reflection of coronal waves through the transition region, and consequent footpoint leakage, are 1-D (e.g. Hollweg 1984, for an early analytical work and Ofman 2002, for a simulation in cold MHD), thus the loop curvature and the possible variation of Alfvén velocity along the loop itself are neglected. Both ingredients are likely to be very important, since the predicted wavelength of kink modes often exceeds the loop curvature radius while a varying Alfvén velocity may change the propagation properties, as we shall see below. Multidimensional simulations have been proposed so far for Alfvénic pulses in *open* coronal structures (Cargill et al. 1997), for fast magnetosonic MHD modes in coronal arcades (Oliver et al. 1998; Arregui et al. 2001), and for standing waves in arcades and loops (Terradas & Ofman 2004). In addition to the above studies, numerical MHD simulations of the propagation of Alfvénic pulses in a compressible plasma, aimed at studying the coupling to magnetosonic modes and phase mixing effects, have been extensively performed (e.g. Malara et al. 1996; Nakariakov et al. 1997; Tsiklauri et al. 2001, 2002a, 2002b, 2003; Selwa et al. 2004). In these works the geometry is simplified by considering constant magnetic fields and density gradients perpendicular to the fieldlines, in order to model just the region near the loop boundaries, thus still neglecting the loop curvature and the stratification by gravity.

In the present paper we make an effort to extend these previous numerical works in two directions. First, we consider transverse oscillations due to the propagation of Alfvénic pulses in a 2-D magnetic coronal arcade, given as a potential field configuration, that models a system of neighboring loops. Thus we do not consider here the usual picture of dense loops in flux tubes, which may act as wave-guide for fast (kink) modes. Second, we include a realistic solar atmosphere, with a model transition region, separating the hot corona from the cooler photosphere, and the density stratification due to gravity. The chosen settings will thus allow us to consider either the leakage at the transition region and the effects on wave propagation and decay in a 2-D inhomogeneous medium, namely WKB losses and the spreading, basically an effective *dispersion*, of the initial pulse along the loop, both consequences of the variation of the background density and Alfvén velocity. Moreover, compressible effects and non-linearity are fully retained, so that mode coupling and wave steepening are allowed in the simulations. As the present paper was in the submission process, we became aware of a paper on 3-D MHD numerical simulations with settings similar to ours which has been published very recently (Miyagoshi et al. 2004). Comparisons and comments will be given in the conclusions.

The paper is structured as follows. In Sect. 2 the initial settings and the simulation setup are described. A brief analytical introduction is given in Sect. 3, whereas Sect. 4 and relative subsections are devoted to the results, for the two cases of symmetric and asymmetric pulses. Further comments on the model and final conclusions are given in Sect. 5.

## 2. Initial conditions and numerical setup

The present work is devoted to the study of the dynamical properties of ideal Alfvénic pulses propagating in a stratified and magnetically structured solar atmosphere. This is a challenging problem even from a numerical point of view, due to the large differences in the characteristic time and length scales imposed by the vertical stratification and by the presence of strong magnetic fields. Since we are mainly interested in the dynamics, the first simplifying assumption that we make is to consider the *ideal* MHD equations, thus neglecting explicit dissipative terms. Moreover we will assume an adiabatic index $\gamma = 1$, so that the (isothermal) energy equation automatically reduces to a continuity-like equation for the pressure. In other words, the temperature jump from the photosphere to the corona is supposed to be maintained by an efficient coupling between heating (in the corona) and conductive and radiative losses (in the transition region and chromosphere), although those terms are not included explicitly in our analysis.

The ideal MHD equations considered in the present paper, written in conservative form (as needed by our shock-capturing method) and in the Gaussian (CGS) system, are:

$$\partial_t \rho + \nabla \cdot (\rho \mathbf{v}) = 0, \tag{1}$$

$$\partial_t p + \nabla \cdot (p \mathbf{v}) = 0, \tag{2}$$

$$\partial_t (\rho \mathbf{v}) + \nabla \cdot [\rho \mathbf{v}\mathbf{v} - \mathbf{B}\mathbf{B}/4\pi + (p + B^2/8\pi)\mathbf{I})] = \rho \mathbf{g}, \tag{3}$$

$$\partial_t \mathbf{B} - \nabla \times (\mathbf{v} \times \mathbf{B}) = 0, \tag{4}$$

$$\nabla \cdot \mathbf{B} = 0, \tag{5}$$

where $\mathbf{g}$ is the surface gravity acceleration, $\mathbf{I}$ is the diagonal identity tensor, and the other quantities have their usual meaning. Notice that the above system is fully nonlinear, plasma compressibility is retained and the temperature, here derived from the equation of state for a fully ionized gas $p = 2nk_BT$ ($n \simeq \rho/m_p$ is the electron and proton number density), is not bound to be constant in spite of $\gamma = 1$, as the density and the pressure are evolved in time separately.

The initial static equilibrium configuration consists of a potential magnetic field, taken as a model for a system of coronal loops (an arcade), embedded in a vertically stratified atmosphere. Let us consider a square computational domain in 2-D Cartesian geometry, with horizontal coordinate $-L/2 < x < L/2$ and vertical coordinate $0 < z < L$, where $L$ is both the arcade width and the computational box size, with $z = 0$ indicating the photospheric level. Translational symmetry is assumed in the other horizontal direction $y$, though the $y$ components of the velocity and magnetic field vectors are allowed (we are assuming the so-called 2.5-D approximation here). In particular $v_y$ will be given as the initial perturbation that drives the subsequent evolution.

To model the coronal arcade we choose a standard potential field solution of the static 2-D MHD equations (e.g. Priest, 1982):

$$B_x = B_0 \cos(kx) \exp(-kz), \quad B_z = -B_0 \sin(kx) \exp(-kz), \tag{6}$$

where $B_0$ is the photospheric field magnitude at the footpoints $x = \pm L/2$ and $k = \pi/L$. Note that the divergence-free condition Eq. (5) is satisfied (the field may be derived from the magnetic potential $A_y(x, z) = (B_0/k) \cos(kx) \exp(-kz)$) and the solution is current-free ($4\pi J_y/c = -\nabla^2 A_y = 0$), as expected for 2-D potential fields. The choice of the above force-free field configuration allows us to decouple the balance of the other forces acting on the system from the magnetic ones. In fact, the divergence of the magnetic part of the momentum tensor in Eq. (3) reduces to the usual form of the Lorentz force $-\mathbf{J} \times \mathbf{B}/c$, which is zero in our settings, so that the pressure gradients must simply balance gravity, directed in the vertical (negative) direction alone. The resulting equilibrium is thus defined by the $z$ component of the momentum equation, which simply becomes

$$\frac{\mathrm{d}p}{\mathrm{d}z} + \rho g = 0, \tag{7}$$

and by using the equation of state we may finally solve for the thermal pressure, once the temperature $T = T(z)$ is prescribed:

$$p(z) = p(z_0) \exp\left[-\frac{m_p g}{2k_B} \int_{z_0}^z \frac{\mathrm{d}z'}{T(z')}\right]. \tag{8}$$

As in Cargill et al. (1997), the temperature profile is taken here as a smoothed (to reduce numerical diffusion) step function, with a low photospheric temperature $T_{\mathrm{phot}}$ and a higher coronal temperature $T_{\mathrm{cor}}$ separated at $z_t$ by a transition region of arbitrary width $z_w$. By using their function

$$T(z) = \frac{1}{2}(T_{\mathrm{cor}} + T_{\mathrm{phot}}) + \frac{1}{2}(T_{\mathrm{cor}} - T_{\mathrm{phot}}) \tanh\left(\frac{z - z_t}{z_w}\right) \tag{9}$$

pressure and density $z$ profiles can be easily integrated analytically.

The thermodynamical quantities and those related to the magnetic field are plotted respectively in Fig. 1 and Fig. 2, for the numerical parameters chosen in the initial equilibrium described above. The unit of length is $10^3$ km $= 10^8$ cm (or 1 Mm), and we assume an arcade width of $L = 50 \times 10^3$ km, velocities will be expressed in units of $10^3$ km s$^{-1}$ = $10^8$ cm s$^{-1}$, proper for the coronal environment, and times in seconds. The photosphere is modeled as a strongly stratified atmosphere with temperature $T_{\mathrm{phot}} = 6000$ K (with a corresponding density scale height of $H_{\mathrm{phot}} = 2k_B T_{\mathrm{phot}}/m_p g \simeq 360$ km, where $g = 2.74 \times 10^4$ cm s$^{-2}$) up to the transition region, placed at $z_t = 2 \times 10^3$ km, where a temperature jump as high as 200 yields a coronal temperature of $T_{\mathrm{cor}} = 1.2 \times 10^6$ K (with a corresponding scale height $H_{\mathrm{cor}} \simeq 72 \times 10^3$ km). Density and pressure are normalized by requiring the value $\rho = 10^{-15}$ g cm$^{-3}$ ($n \simeq 6 \times 10^8$ cm$^{-3}$) at the coronal base, just above the transition region. The magnetic field is normalized to $B_0 = 40$ G, which leads to a maximum of the Alfvén speed $v_A$ of $\simeq 3000$ km s$^{-1}$ (at the coronal base) and to a corresponding minimum value of the plasma beta, defined here as the squared ratio of the sound speed to the Alfvén speed, of about $2 \times 10^{-3}$.

The large density jump of four orders of magnitude from the photospheric base to the corona, and the correspondent jump of a factor 100 in the Alfvén speed and thus in the wave propagation characteristic time scales, do not allow to follow numerically the dynamics in both environments. Therefore, in the present simulations only the (fast) coronal dynamics will be studied with sufficient temporal and spatial resolution, whereas waves penetrating the atmospheric layers below will appear as almost steady, poorly resolved features.

Notice that since the field magnitude $B = |\mathbf{B}|$ depends on $z$ alone, as $B = B_0 \exp(-kz)$ from Eqs. (6), also the Alfvén speed (and the plasma beta) will depend on $z$ only. In particular, in the corona we can write

$$v_A = B (4\pi\rho)^{-1/2} \propto \exp[-(1 - \delta/2) kz], \tag{10}$$

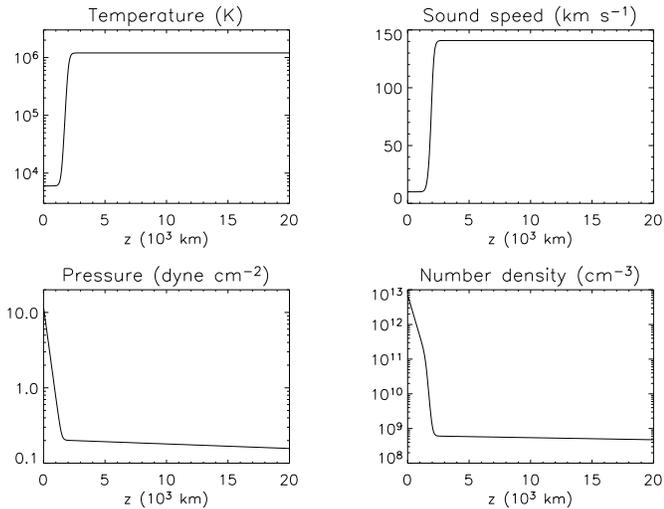

**Fig. 1.** The initial configuration: thermodynamical quantities. The temperature jump at the transition region is 200, ranging from $T_{\rm phot} = 6000$ K to $T_{\rm cor} = 1.2 \times 10^6$ K. Note that less than half of the entire $z$ domain is displayed, to better appreciate the gradients in the photosphere and at the transition region.

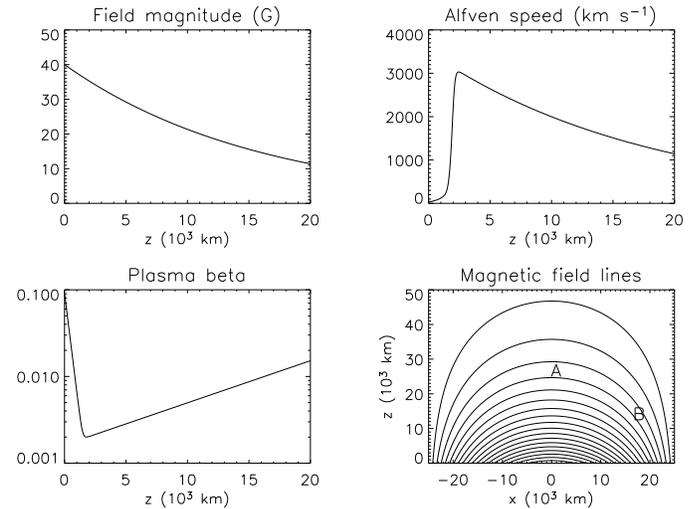

**Fig. 2.** The initial configuration: magnetic field and related quantities. Again the vertical extent of the domain is not entirely plotted. In the last plot the two positions of the initial Alfvénic pulse (given as a pulse-like perturbation in the transverse velocity) are also indicated (run A: symmetric case, run B: asymmetric case).

where

$$\delta = \frac{1}{kH_{\rm cor}} = \frac{L/\pi}{H_{\rm cor}} \qquad (11)$$

is the ratio between the magnetic and density scale heights (defined also in Oliver et al., 1998). Since the background coronal Alfvénic velocity will be the main driver for the perturbations evolution and propagation, its functional profile in Eq. (10), determined as we have just seen by the combined assumptions of a potential field magnetic arcade and of a nearly isothermal corona, is a fundamental issue. When $\delta < 2$, as in our case ($\delta \simeq 0.22$, by using the realistic numbers given above), the coronal Alfvén speed is a decreasing function of height (while the beta increases), the opposite holding when $\delta > 2$. A constant coronal Alfvénic velocity profile is found for $\delta = 2 \Rightarrow L = 2\pi H_{\rm cor}$.

The perturbation at $t = 0$ is taken as a purely transverse velocity pulse $v_y$, with $B_y = 0$, of normalized amplitude $A$ and located at $(x_0, z_0)$:

$$v_y = \frac{A v_0}{1 + (r/r_0)^4}, \qquad (12)$$

where $r = [(x - x_0)^2 + (z - z_0)^2]^{1/2}$, $r_0 = 10^3$ km, and $v_0 = 10^3$ km s$^{-1}$, which is the typical Alfvén speed in the corona (reached at $z \simeq 22.4 \times 10^3$ km in our settings). In Fig. 2 the two locations of the initial pulse considered here are also shown: the symmetric case (run A) refers to a pulse located at the center of the computational domain ($x_0 = 0, z_0 = L/2$), whereas the asymmetric case (run B) refers to a pulse located at ($x_0 = 2/3 L/2, z_0 = L/4$), approximately along the same fieldlines. The two cases considered here simulate the response of a coronal loop which is reached by blast waves produced by nearby flares, case A for perturbations at the top of the loop and case B for perturbations along one of its legs. In Sect. 4 we shall describe separately the results of the simulations in both cases, for different values of the normalized amplitude $A$.

Numerical simulations will be here performed with a third order (both in space and time) shock-capturing scheme based on an ENO-type (*Essentially Non Oscillatory*, see Shu 1997, for a review) finite-difference scheme, with a simple Lax-Friedrichs solver used in upwind fluxes. Full description of the code and of a general method to preserve numerically the divergence-free constraint Eq. (5) in shock-capturing schemes (*Upwind Constrained Transport*, UCT) may be found in Londrillo & Del Zanna (2000, 2004). The method and the code have been recently extended to relativistic MHD for high energy astrophysics applications (Del Zanna et al. 2003).

Due to the large gradients involved at the transition region and to the presence of curved magnetic fields in the low-beta corona, numerical diffusion of the equilibrium quantities is a serious issue, especially for conservative schemes, and some *ad hoc* treatments must be adopted. Our choice is to subtract from the numerical fluxes reconstructed at intercells the contributions involving the equilibrium quantities alone, retaining of course full non-linearity. In this way the equilibrium is stable by definition, regardless of the resolution employed. Wave propagation is instead affected by resolution. We find that $400 \times 400$ grid points, with uniform spacing in both directions, is a good compromise between precision and efficiency. Runs at lower resolution ($200 \times 200$) already give the correct picture, though numerical diffusion is still apparent because the pulse is sampled with a small number of grid points. On the other hand, higher resolution ($600 \times 600$) does not improve significantly the results in comparison with the adopted one (e.g., in Fig. 4 discrepancies are less than 1% when motions affect the corona alone, somehow higher after the interaction with the steep transition region, since if it is moved from its original position diffusion sets in).

Finally, characteristics-based (transparent) outflow boundary conditions (see e.g. the appendix in Del Zanna et al. 2001) are applied at $z = 0$ and $z = L$, whereas reflecting conditions are imposed at $x = \pm L/2$, in order to prevent the flow from crossing the $x$ boundaries ($v_x$ vanishes automatically there), simulating in this way a row of arcades in the $x$ direction.

## 3. Basics of pulse propagation

In the linear 1-D MHD case, a velocity perturbation normal to a background magnetic field in a uniform plasma is known to split into two Alfvén waves propagating in opposite directions, each of them preserving the original shape but with half the initial amplitude. The wave moving in the positive direction will be characterized by an anti-correlated transverse field perturbation $\delta \boldsymbol{B} = -(4\pi\rho)^{-1/2}\delta\boldsymbol{v}$, while positive correlation will hold for the wave moving in the negative direction. The ponderomotive force due to the unbalanced magnetic field, with a quadratic (second order) dependence on the pulse amplitude, triggers compressible pressure and density fluctuations which then propagate as fast and slow modes.

In our 2-D setup, the propagation of the MHD modes will depend also on the shape of the underlying coronal field and density gradient, with Alfvén and slow modes following the curved path determined by the potential arcade, while the fast mode will basically propagate isotropically. The speed of propagation is of course expected to vary and both the amplitude and shape of the pulses will change during propagation.

For example, consider the 2-D propagation of the two Alfvén waves triggered by our initial pulse. By using the Elsässer formalism, the linearized equations neglecting mode coupling and reflection are (e.g. Velli 1993):

$$\partial_t z^\pm \pm \boldsymbol{v}_A \cdot \nabla z^\pm \mp \frac{1}{2}(\nabla \cdot \boldsymbol{v}_A) z^\pm = 0, \quad (13)$$

where $\boldsymbol{x} \equiv (x, z)$ and $z^\pm(\boldsymbol{x}, t) = v_y(\boldsymbol{x}, t) \mp B_y(\boldsymbol{x}, t)/\sqrt{4\pi\rho(\boldsymbol{x})}$ are the two pulses propagating respectively parallel or antiparallel to the local background $B(\boldsymbol{x})$. Since $\nabla \cdot \boldsymbol{v}_A = -\frac{1}{2}\boldsymbol{v}_A \cdot \nabla\rho/\rho$, it is convenient to solve for the normalized quantities $z^\pm(\boldsymbol{x}, t) = [\rho(\boldsymbol{x})]^{-1/4} f^\pm(\boldsymbol{x}, t)$, so that the equations for $f^\pm$ simply become

$$[\partial_t \pm \boldsymbol{v}_A \cdot \nabla] f^\pm(\boldsymbol{x}, t) = 0. \quad (14)$$

The above wave equations are now in the standard first-order form and may be solved by the method of characteristics (or *ray-tracing* techniques). Thus, the following modifications are expected to arise with respect to the 1-D case:

1. Wave propagation now occurs along curved paths (the coronal loop).
2. The background Alfvén speed changes along the loop and this leads to both a varying group velocity and to a deformation of the initial shape, since different portions of the wave packet will experience different phase speeds. In our particular case we will see an *acceleration* and a *spreading* (basically a dispersion-like effect) of the pulses when moving towards the transition region, where the Alfvén velocity is higher.
3. The amplitude changes systematically as $\rho^{-1/4}$ and $\rho(\boldsymbol{x}) \equiv \rho(z)$ is a decreasing function of height, so the pulses amplitude will decrease while propagating downwards (WKB losses), producing the apparent *damping* (the energy flux is actually conserved).

Concerning slow and fast modes, since the initial perturbation is in our case transverse to *both* the background magnetic field and the wave vector plane, that is $\nabla \cdot \boldsymbol{v} = 0$ at $t = 0$, thus coupling to magnetosonic modes is still expected to be due to the ponderomotive force alone and hence to depend quadratically on the initial $v_y$ amplitude.

All the above analytical predictions will be confirmed by our numerical simulations, as discussed in the following section.

## 4. Numerical results

### 4.1. Run A: symmetric case

For the first simulation in the symmetric case we take a normalized transverse velocity amplitude $A = 0.1$ in Eq. (12), which corresponds to 100 km s$^{-1}$ (the background Alfvén speed is slightly less than $v_0 = 10^3$ km s$^{-1}$ at the pulse initial position $z = L/2$).

In Fig. 3 the time evolution of the transverse $v_y$ velocity component (upper row) and of density perturbations (lower row), normalized to the equilibrium density, is shown for the entire $x - z$ domain. We can follow the propagation of the various MHD modes as described in Sect. 3 in the simulations: the twin Alfvénic pulses, each of initial amplitude of $v_y = (A/2)v_0$, move downward following the same fieldline on which they had been originally placed (the plot at $t = 10$ s). At the same time, the associated compressible fluctuations propagate as fast modes (the external arc) and slow modes (the inner couple of fluctuations). Notice the absence of trapped fast modes in the loop, since we do not have a higher density structure which may act as a wave-guide for such modes. As expected from the analysis in Sect. 3, the transverse quantities suffer WKB losses due to the increase of the background density, and at the same time spread and deform along their path because of the increase in the local Alfvén speed. For similar reasons fast waves present a narrow, compact front when propagating upward, while the downward propagating front is hardly visible, since it soon spreads at always increasing speed. Slow modes do not show this behavior, and steadily propagate (along $\boldsymbol{B}_0$) at approximately the sound speed $c_s \simeq 140$ km s$^{-1}$.

At later times, say $t = 40$ s, the transverse waves have already bounced at the transition region (where they have reverted sign), reaching approximately the same positions as at $t = 10$ s. The waves are now propagating in regions with a lower density, so the front of each pulse assumes a more compact shape. Time $t = 100$ s is about the total crossing time (the pulses period), since each pulse has hit the transition twice and converged back to the center of the computational domain. Note the effect of the spreading, which is even more evident at time $t = 200$ s, that goes together with a huge damping in the amplitude of the waves (about a factor of three after each bounce). This is not only due to WKB losses but also to reflection at the transition region, where the profile of the pulses is somehow convolved with that of the density, producing an additional dispersion of the wave packet.

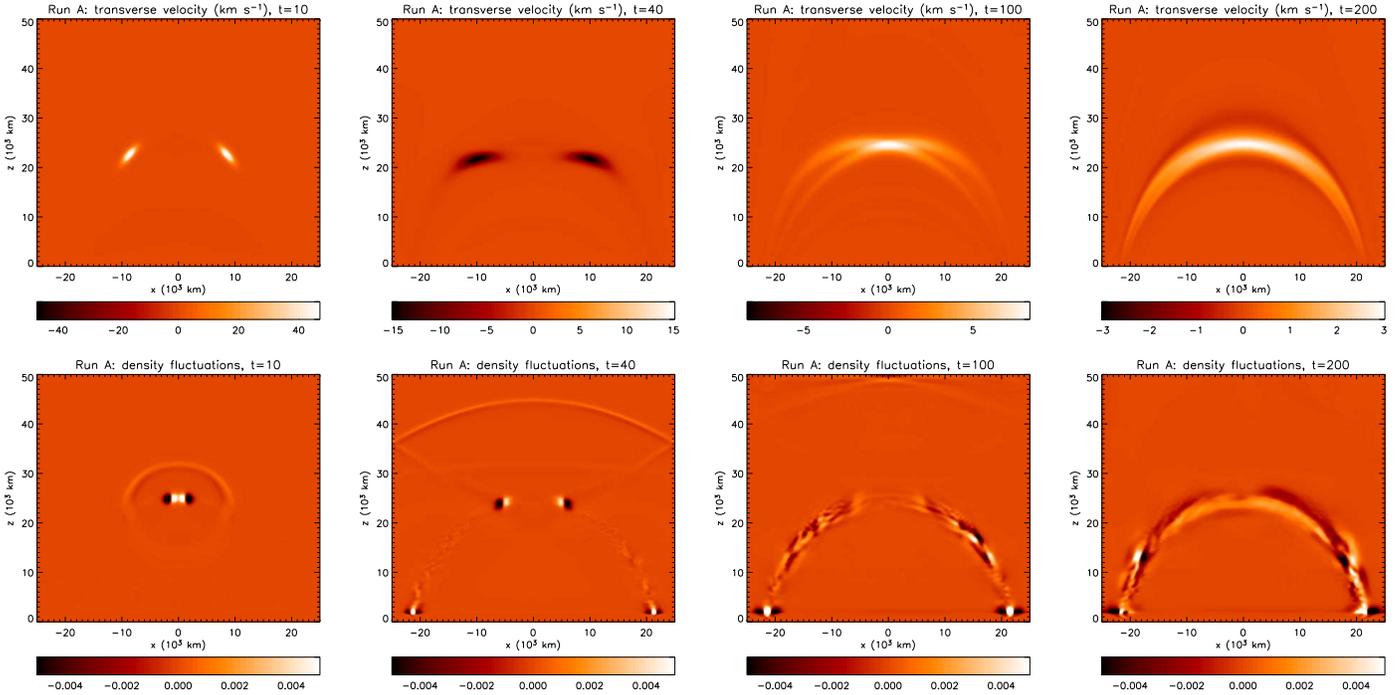

**Fig. 3.** Run A (symmetric case): transverse velocity (upper row) and relative density perturbations (lower row) at times $t = 10$ s, $t = 40$ s, $t = 100$ s, and $t = 200$ s. Note that the color tables refer to symmetric ranges whose bounds are defined as $\pm\max(|v_y|)$ in the upper row (decreasing with time) and as 0.005 in the lower row (kept constant in time).

As far as compressive fluctuations are concerned, in the $t = 40$ s plot the fast waves reflected off the lateral boundaries are clearly visible (the correspondent amplitude is however very small), as well as the fast wave front which has been reflected off the transition region itself (with a larger curvature radius, at a height of $\simeq 30 \times 10^3$ km), which is just about to follow the first front in leaving the domain at $t = 100$ s. More important are the density perturbations induced at the transition region by the Alfvén waves hitting the density wall. These perturbations may be quite large (up to $\lesssim 0.01$, saturated in our plots), since it is the transition region itself which is displaced from the original position. The compressed plasma drives slow waves, which appear as standing oscillations along the transition region, flow back into the corona along the arcade legs and eventually penetrate the photosphere (at about 10 km s$^{-1}$, the local sound speed, too slow to be followed in our simulations) toward the arcade footpoints. Finally, driven magnetosonic fluctuations are also produced nonlinearly by the traveling Alfvén waves, apparently *dragged* along the arcade by the mother waves and soon filling the entire path with compressive noise.

The presence in our simulations of compressive modes associated to the propagation of the Alfénic pulses may also provide a common theoretical background for both the observations of the fast transverse oscillations and the slow-mode longitudinal waves (see the Introduction).

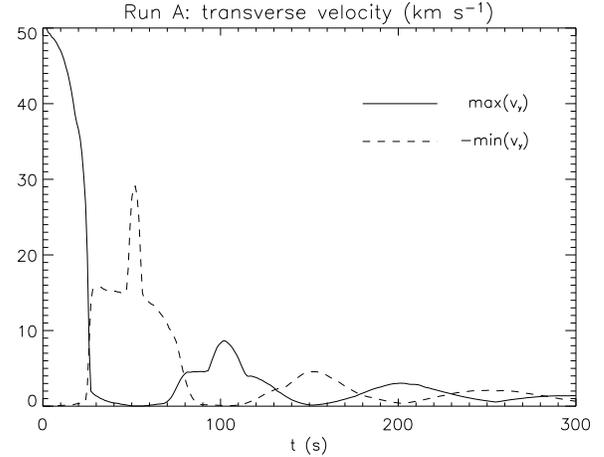

**Fig. 4.** Run A (symmetric case): the transverse velocity as a function of time. The maximum (solid line) and minus the minimum of $v_y$, that is the pulses peak values, are plotted in the figure.

The time evolution up to $t = 300$ s of the transverse velocity component is summarized in Fig. 4, where the maximum and minimum (with reverted sign) over the whole domains are computed (rms quantities show similar behaviors). The peaks occurring every $\sim 50$ s are due to the superposition of the two waves (including the initial value $v_y = 100$ km s$^{-1}$), whereas the strong damping caused by the stretching of the pulses and by the interaction with the transition region at every bounce (leakage to the photosphere, coupling to the transition region finite width, mode conversion to compressible fluctuations) is apparent. It is unfortunately very hard to disentangle these combined effects to provide quantitative measurements of each of those, also because of a loss of accuracy of the numerical scheme (due to the large gradients involved and to additional numerical diffusion) when the transition region is displaced from its initial position, and because of the poor resolution in the photospheric region. A rough estimate of the fraction of the Alfvénic energy that leaks into the photosphere however gives a figure around 10%, while that going into magnetosonic waves seems to be less than 1%.

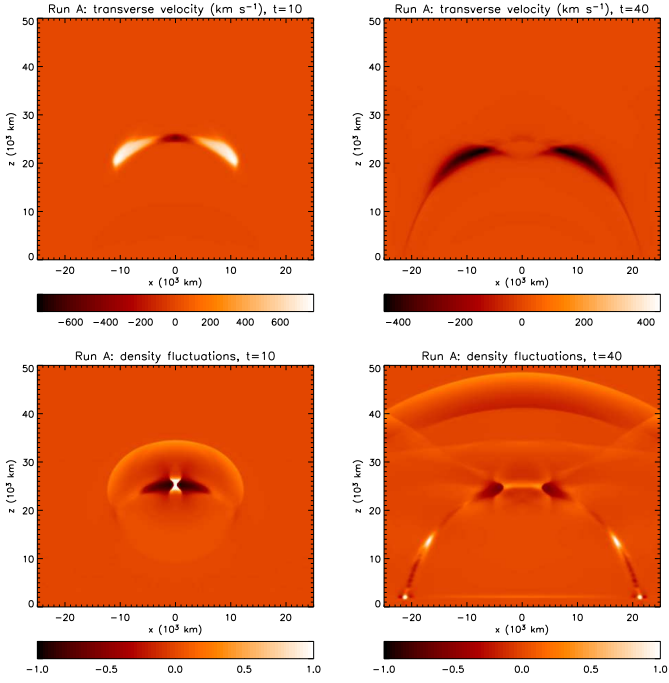

**Fig. 5.** Run A (symmetric case): transverse velocity (upper row) and relative density perturbations (lower row) at times $t = 10$ s and $t = 40$ s in the high amplitude ($A = 5$) case.

The waves treated so far are basically linear, since even at the initial time each of the opposite propagating velocity pulses has a peak of $A/2 \simeq 5\%$ with respect to the local Alfvén speed. Runs with higher amplitudes have been performed and the results are qualitatively the same as long as the initial amplitude is less than the background Alfvén speed. In particular it has been possible to verify that the slow and fast modes generated initially by the ponderomotive force scale exactly as $A^2$, as expected.

Let us now consider a highly nonlinear case by taking $A = 5$, that is an initial velocity of 5000 km s$^{-1}$. The results before and after the first bounce are reported in Fig. 5, at times $t = 10$ s and $t = 40$ s as in the first two columns of Fig. 3. By comparing to the previous case, the nonlinear effects at work on the profile of the Alfvénic pulses are apparent. These have steepened in the direction of the motion and thus present a highly non-symmetrical shape even at very early times. Correspondingly the compressive fluctuations, now even of order one, look highly distorted too. In spite of these differences the plot of the decaying wave amplitudes as a function of time looks very similar to that shown for the linear case, ob-viously with values proportional to $A$, and we do not show it here.

### 4.2. Run B: asymmetric case

Let us now consider the asymmetric case, that is when a transverse perturbation is induced by a nearby flare not at the top of the arcade but along one of its legs (the assumed reflecting boundary conditions at $x = L/2$ assure that the arcade centered at $x = L$ is perturbed as well in a symmetric fashion).

The results are shown in Fig. 6, where the same outputs as in the previous section have been reported. Differences in the Alfvénic and compressible fluctuations are apparent, since symmetry is clearly lost. At $t = 10$ s, due to the higher local Alfvén velocity in the lower part of the corona (almost 2000 km s$^{-1}$ at the initial height $z = L/4$), the downward propagating wave has already bounced off the transition region and reverted its sign in $v_y$. Both pulses then move toward the left footpoint of the arcade and at $t = 40$ s they appear to converge there together. Also at later times there is the contemporary presence of positive and negative components, which have spread along the whole arcade, as in the previous case. Compressive fluctuations show the usual fast wave fronts and slow mode peaks moving along the fieldlines. At $t = 40$ s also the reflected fast wave front is visible. The density perturbations created at the transition region due to its displacement from the original position and the associated slow modes are now larger than in the previous case, possibly because the pulse is less stretched when it hits the density gradient (we maintain, however, the same range in the color plots to enhance the coronal density fluctuations). The situation at later times is basically the same as in run A.

In Fig. 7 we plot the time evolution of max($v_y$) and $-$min($v_y$), calculated over the whole domain, as in the previous section. Note that the damping is even stronger than in case A, and Alfvénic pulses never combine together in a constructive way; on the contrary, pulses with opposite signs are always present together at all times. The present situation may resemble the scenario described by Aschwanden et al. (2002), where in many different loops transverse pulses were observed to propagate non-symmetrically back and forth along the fieldlines and extremely short damping times, basically coincident to the crossing times, were estimated.

We do not report here the high-amplitude case for run B, for which the same kind of differences with respect to the linear case shown in the previous section are found.

## 5. Discussion and conclusions

The 2.5-D MHD simulations presented in this paper are aimed to include the effects of a stratified and inhomogeneous atmosphere in the propagation of Alfvénic pulses. We have demonstrated that a simple explanation for the observed fast damping of transverse oscillations in coronal loops might be provided by the *dispersion* and amplitude decay of such pulses induced by background density and magnetic field gradients along the loop.

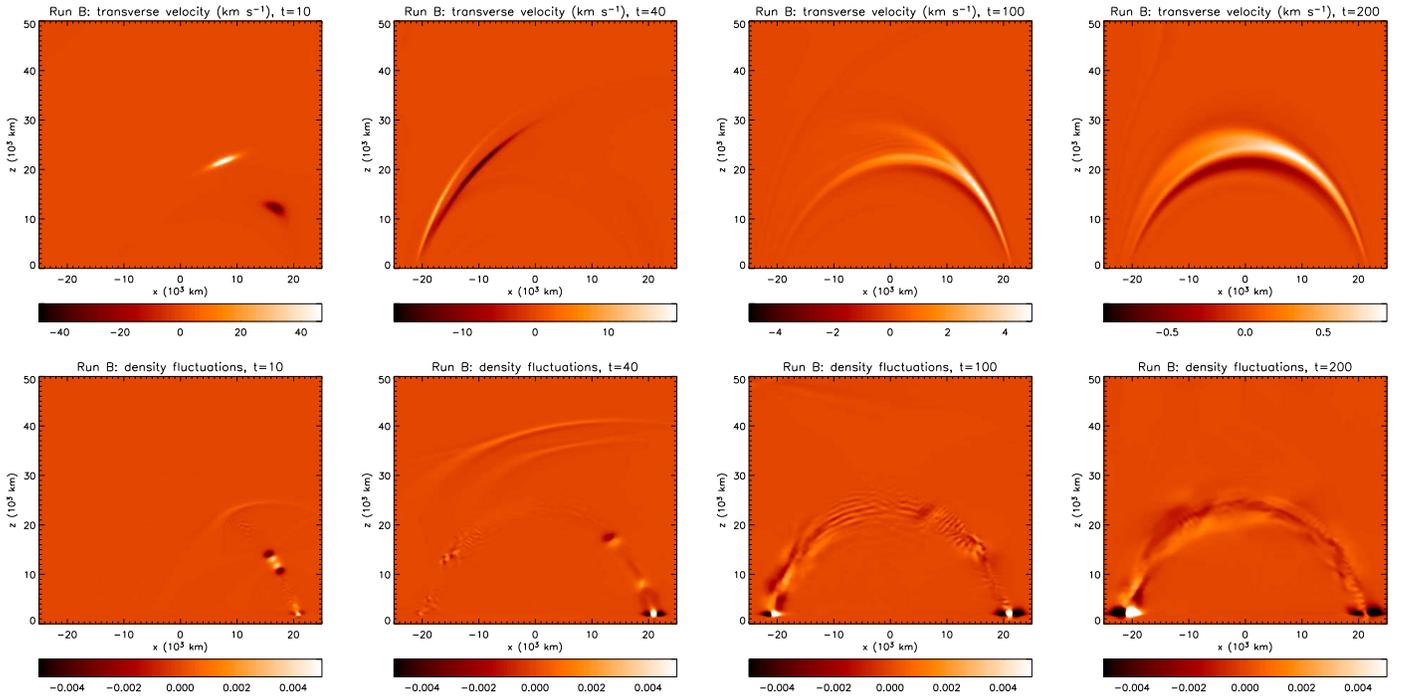

**Fig. 6.** Run B (asymmetric case): transverse velocity (upper row) and relative density perturbations (lower row) at times $t = 10$ s, $t = 40$ s, $t = 100$ s, and $t = 200$ s. Note that the color tables refer to symmetric ranges whose bounds are defined as $\pm|\max(v_y)|$ in the upper row (decreasing with time) and as 0.005 in the lower row (kept constant in time).

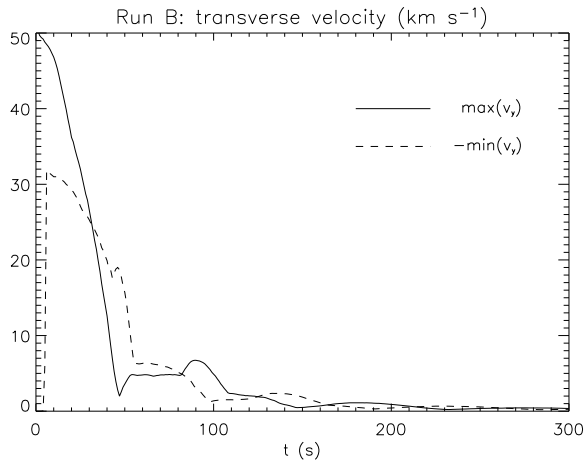

**Fig. 7.** Run B (asymmetric case): the transverse velocity as a function of time. The maximum (solid line) and minus the minimum of $v_y$, that is the pulses peak values, are plotted in the figure.

Coronal loops are usually modeled as elongated cylindrical structures with uniform density and magnetic field, where these quantities are allowed to change in the transverse direction alone, across the loop boundaries (flux-tube models). Our approach is different. We have considered a 2-D arcade as a system of neighboring loops in a stratified corona, where all quantities vary also *along* the loop. In this case we have wave propagation in a highly inhomogeneous medium: WKB losses in the wave amplitude are at work when the pulses are moving downward in the corona, where the density increases, and at the same time the pulses suffer a strong spreading, due to the increasing background Alfvén speed. Additional stretching and damping arise when the pulses interact and are reflected off the transition region, also partly due to leakage to the photosphere and to the coupling with compressive modes. What is missing in our model is the density enhancement along the loop and thus the possibility of trapping the fast modes like in flux-tube models.

How do our simulations relate to the observations? As discussed in the Introduction, it appears that the flare-generated transverse oscillations observed in coronal loops may be vaguely classified into two categories. In the first one are those oscillations which are generally identified as global kink-type standing modes. To this class belong the few known examples where the loop magnetic field strength has been tentatively inferred by applying the theory of MHD modes in cylindrical flux-tubes (Nakariakov & Ofman 2001). In the second class lies the vast majority of the reported observations (Aschwanden et al. 2002), where the oscillations seem to be rather due to a superposition of impulsively generated MHD waves which propagate back and forth and quickly decay (in $1-3$ Alfvénic crossing times).

Our simulations are an attempt to model the second class of observed cases, thus we have chosen a *localized* initial disturbance, either centered or not in our system of loops, rather than imposing an initial motion to the whole loop. The results seem to confirm the scenario described in the above paper, since the pulses move back and forth along the loop, initially as a couple of elongated features and soon as a superposition of modes with different signs and direction of propagation. After a few crossing times nearby fieldlines may appear to oscillate almost

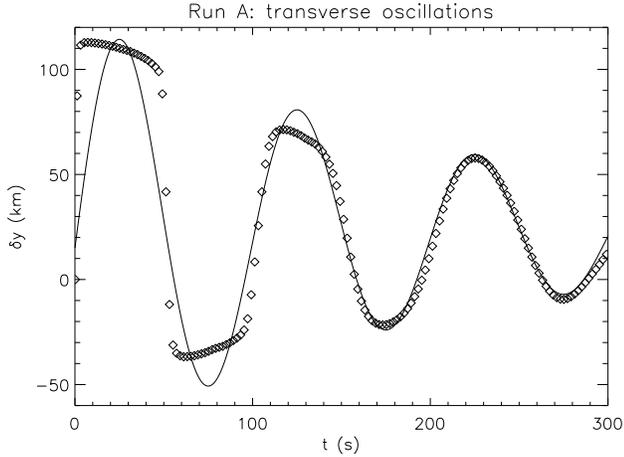

**Fig. 8.** Run A (symmetric case): the transverse displacement, calculated by integrating $v_y$ at the center of the computational domain (the site of the disturbance at $t = 0$) as a function of time (symbols) and its fit by using Eq. (15) (solid line). The ratio between the resulting damping time and the period is $\tau_d/P \simeq 2.2$.

as global Alfvénic modes, in the sense that the single loops move in phase all along their length.

In order to attempt a match of the decay times found in our simulations with that measured by observing the transverse displacement of real coronal loops, we have integrated in time the $v_y$ velocity component of the pulses *at a fixed position along the loop*, namely the site of the pulse at the beginning of the simulation. The profile of the displacement as a function of time is plotted in Fig. 8 for the symmetric case A. The solid line refers to a fit made by assuming a damped cosine function for the velocity,

$$\delta y(t) = s_0 + \int_0^t C \cos\left(2\pi t'/P\right) \exp\left(-t'/\tau_d\right) dt', \quad (15)$$

with $s_0 = 15$ km, velocity amplitude $C = 6.5$ km s$^{-1}$, wave period $P = 100$ s, and damping time $\tau_d = 220$ s. As we can see, in spite of the initial differences due to the rapid transits of the pulses at the considered site, after one or two periods the simulated and fit functions become almost identical. The ratio between the damping time and the period is $\tau_d/P \simeq 2.2$ in this case (even smaller ratios are found in the asymmetric case, not shown here), which is well in the range of the data set by Aschwanden et al. (2002; Fig. 18). The other values of the fit (oscillation amplitude, period, decay time) are all in the lower bounds of the data set, but we should remember that we had chosen a small loop (correspondent to an arcade width of $L = 50 \times 10^3$ km) for numerical convenience, and the Alfvénic crossing time obviously scales linearly with this quantity.

Comparing our work with the recent paper by Miyagoshi et al. (2004), we find they use very similar initial settings as in our case (a 2-D potential field arcade embedded in a vertically stratified atmosphere), but their initial $v_y$ perturbation also depends on $y$, thus introducing 3-D effects in the subsequent evolution. The velocity at the site of the initial perturbation shows a similar time dependency as we have found, though we do not find fully convincing the explanation provided for the damping, namely the leakage of the energy of fast modes out of the simulation box. The mechanism described in the present paper applies basically unchanged to the above 3-D settings, so this should actually determine the observed decay. Moreover, due to the large gradients involved and to the use of a second order scheme, the numerical resolution employed by Miyagoshi et al. is probably too low.

The mechanism suggested here for the observed fast decay of transverse oscillations obviously does not rule out the other explanations referenced in the Introduction, though we have demonstrated how crucial is the inclusion in the modeling of the loop curvature and of a varying Alfvén velocity. The curvature should also be included in flux-tube models, since global trapped kink modes have a wavelength of the same order of the length of the loop and hence of the curvature radius itself, thus clearly 2-D effects must be important.

The preliminary simulations presented in this paper should be extended by exploring a wider range of parameters. For example the width of the initial pulse is a key ingredient, since global oscillations may be more efficiently produced for larger pulses. Moreover, our model could provide a diagnostic tool to infer important properties of the loops, in the spirit of coronal seismology. For example, the damping rates obviously depend on the assumed Alfvén speed profile (namely our parameter $\delta$). Another aspect which has not been investigated is the possible link between the slow magnetosonic modes which are triggered by the Alfvénic pulses in our simulations and the observations of such modes by the SOHO and TRACE instruments, as mentioned in the Introduction. We leave these tasks as future work.

*Acknowledgements.* The authors thank C. Chiuderi, J. Kuijpers, V. Nakariakov, S. Landi, G. Del Zanna for stimulating discussions, and an anonymous referee for useful comments that helped to improve the paper. This work was supported by the Italian MIUR grant Cofin 2002 (code 2002-02-5872) and by the European Union RTN (code HPRN-CT-2001-00310) and Erasmus programs.